\documentclass[a4paper,fleqn]{cas-sc}

\usepackage[natbibapa]{apacite}
\def\tsc#1{\csdef{#1}{\textsc{\lowercase{#1}}\xspace}}
\tsc{WGM}
\tsc{QE}
\tsc{EP}
\tsc{PMS}
\tsc{BEC}
\tsc{DE}

\begin{document}
\let\WriteBookmarks\relax
\def\floatpagepagefraction{1}
\def\textpagefraction{.001}

\shorttitle{Assessment of Personalized Learning in Immersive and Intelligent Virtual Classroom on Student Engagement}

\author[1]{Ying Weng}[type=editor,orcid=0000-0003-4338-713X]
\cormark[1]

\author[1]{Yiming Zhang}[type=editor,orcid=0000-0002-9570-8962]



\affiliation[1]{
    addressline={School of Computer Science, Faculty of Science and Engineering, University of Nottingham Ningbo China}, 
    city={Ningbo},
    citysep={}, 
    postcode={315100}, 
    state={},
    country={China}}

\title [mode = title]{Assessment of Personalized Learning in Immersive and Intelligent Virtual Classroom on Student Engagement}                      
\cortext[cor1]{Corresponding author}



\begin{abstract}
As trends in education evolve, personalized learning has transformed individual’s engagement with knowledge and skill development. In the digital age, state-of-the-art technologies have been increasingly integrated into classrooms to support intelligent education and foster personalized learning experiences. One promising approach is the use of eye-tracking technology to evaluate student engagement in intelligent virtual classrooms. This paper explores the assessment of personalized learning in the virtual classroom and its impact on student engagement through the eye movement paradigm. The study aims to provide insights into how personalized learning approaches can enhance student participation, motivation, and academic performance in the online learning environment. Through a comprehensive literature review, case study, and data analysis, the paper examines the key elements of personalized learning, the methods of assessment, and the resulting effects on student engagement. The findings suggest that the eye movement paradigm has the potential to assess student engagement and promote better educational outcomes.

\end{abstract}




\maketitle

\section{Introduction}

In the digital age, virtual classrooms have become an increasingly prevalent mode of education. As educational institutions strive to meet the diverse needs of students, personalized learning has emerged as a promising approach. Personalized learning aims to customize education based on students' individual needs, interests and learning styles, making the learning experience more efficient and relevant \citep{shemshack2020systematic}. This mode of teaching not only relies on big data analysis, artificial intelligence and other technologies to identify students' learning habits, but also adjusts the teaching content in real time through the interactive functions of the online platform, helping students to better learn new knowledge and improve their learning results.

Studies have shown that personalized learning can significantly improve students' motivation and performance. For example, a study by \citet{pane2015continued} found that students who adopted personalized learning methods showed significant improvements in math and reading performance over students in traditional teaching models. In addition, \citet{basham2016operationalized} emphasized that personalized learning in virtual learning environments can effectively promote the development of self-regulated learning skills and contribute to students' long-term learning development. In contrast, \citet{roschelle2006co} pointed out that personalized learning combined with cooperative learning can create a more comprehensive learning experience for students and improve learning effectiveness.

Virtual classrooms represent a new paradigm in education, facilitated through the internet. This mode of learning makes it easier for students to participate remotely, and it played a crucial role during the COVID-19 pandemic by enabling the maintenance of social distancing while continuing education. The use of novel techniques including Artificial intelligence (AI) and virtual reality (VR) can make the virtual classroom more intelligent \citep{demszky2024automated, huang2023effects}. Personalized learning, especially within the context of intelligent virtual classrooms, has become increasingly important as it allows tailored educational experiences that cater to individual student needs, thereby enhancing learning outcomes. For example, \citet{huang2020disrupted} emphasized how AI-powered virtual classrooms can adapt content to suit each learner’s pace and preferences, significantly improving engagement and comprehension. Similarly, \citet{anderson2011three} discussed how the flexibility of online learning environments enables the application of personalized learning strategies at scale, making education more accessible and efficient for diverse student populations. VR is an innovative technology that offers an immersive experience \citep{huang2023body, jongbloed2024immersive}. However, its application in education, particularly in virtual classrooms, has not yet been thoroughly explored.

One of the primary advantages of personalized learning in intelligent virtual classrooms is the ability to address the diverse learning needs of students. Every learner has a unique way of absorbing and processing information. Intelligent virtual classrooms can use data-driven insights to identify a student's preferred learning style and present the content accordingly. Personalized learning also enables students to progress at their own pace. In a traditional classroom setting, students are often forced to move along with the majority, regardless of whether they have fully understood the material or not. In an intelligent virtual classroom, students who grasp concepts quickly can move on to more advanced topics, while those who need additional time and support can receive targeted remediation and extra practice. This individualized pacing helps prevent students from falling behind or getting bored due to slow or fast instruction.

Another significant benefit is the enhanced engagement and motivation it offers. When students are presented with content that is relevant and interesting to them, they are more likely to be actively involved in the learning process. Intelligent virtual classrooms can use data on a student's hobbies, interests, and previous academic performance to recommend courses, projects, and activities that align with their passions. However, to understand the effectiveness of personalized learning in the virtual classroom, it is essential to assess its impact on student engagement.

Student engagement is a critical factor in determining the success of educational endeavors. Engaged students are more likely to be actively involved in the learning process, persist in their studies, and achieve better academic outcomes. Therefore, assessing how personalized learning influences student engagement is of paramount importance in optimizing the intelligent virtual learning environment.

\subsection{Research Questions (RQs)}
Despite the growing body of research on personalized learning and virtual classrooms, no studies to date have specifically investigated the use of immersive and intelligent virtual classrooms for personalized learning assessment through the eye movement paradigm. To address this research gap, this study aims to develop and evaluate a virtual reality application designed to assess eye movements in a virtual classroom setting. By doing so, we seek to provide insights into how eye-tracking technologies can be leveraged to enhance the assessment of student engagement and personalized learning experiences. In this paper, we conduct a case study to explore the effectiveness of student engagement in the virtual classroom environment by focusing on the following research questions:

(1) Can 3D eye-tracking paradigms be utilized to assess personalized learning in virtual classrooms and their impact on student engagement?

(2) What are the effective eye-movement metrics for assessing student engagement in a VR-based virtual classroom?

The remainder of the paper is structured as follows: Section 1 provides the introduction and presents the two key research questions that this study aims to address. Section 2 offers a comprehensive literature review, covering recent advancements in both traditional personalized learning and personalized learning within virtual environments. The methodology section outlines the experimental design, while the results and discussion sections present and analyze the experimental findings, and offers recommendations for future research directions. Finally, the conclusion summarizes the key research outcomes. \\ \\

\section{Literature Review}

Research has demonstrated that personalized learning can enhance student motivation, self-efficacy, and the perceived relevance of learning materials. This section reviews the literature on personalized learning and student engagement in both traditional and virtual educational environments.

\subsection{Traditional Personalized Learning}

In traditional personalized learning environments, the focus is on customizing educational experiences to meet the individual needs of each student, typically based on their prior knowledge, learning preferences, and pace. The aim is to create a more engaging and effective learning process by tailoring instruction to the unique characteristics of each learner. \citet{lin2013data} found that when students were given personalized learning paths based on their prior knowledge and skills, they reported higher levels of interest and commitment to the learning tasks. Another study by \citet{alamri2020using} demonstrated that personalized feedback and adaptive instructional strategies improved student engagement and academic performance in online courses.

Conversely, studies on student engagement have identified factors such as interaction with instructors and peers, autonomy in learning, and the quality of instructional design as significant predictors of engagement. Understanding these existing research findings provides a foundation for investigating the specific relationship between personalized learning and student engagement in the virtual classroom context.

\subsection{Personalized Learning in Virtual Environment}

In the virtual classroom, personalized learning can take various forms. Adaptive learning systems use algorithms to adjust the difficulty level and content of lessons based on student responses and performance \citep{ennouamani2017overview}. Intelligent tutoring systems provide individualized feedback and guidance \citep{mousavinasab2021intelligent}. Personalized learning plans can be developed for each student, taking into account their goals, interests, and learning pace.

As stated by \citet{chen2020artificial}, AI has been extensively adopted in educational areas, such as intelligent education systems and humanoid robots, which significantly enhance the quality of learning. Likewise, the examination conducted by \citet{kumar2023exploring} demonstrates that AI technologies could be leveraged to create new possibilities for personalized education and to enhance teaching and learning experiences. According to \citet{katiyar2024ai}, AI techniques including machine learning and natural language processing can be utilized in creating personalized learning experiences. This research highlights that AI-driven personalized systems have the potential to improve student engagement and educational outcomes.

In terms of the integration of Artificial Intelligence and virtual learning environments, \citet{wei2021review} found that intelligent agents in virtual learning environments could utilize procedural memory to respond to the learner's perceptions and take appropriate actions through immediate feedback. This research also demonstrates that as the level of intelligence increases, the disparity between the virtual and real learning environments decreases. As noted by \citet{kavitha2019critical}, AI is used in various stages of the virtual learning process, including diagnosing knowledge levels, analyzing learning engagement, and giving individualized suggestions. In addition, \citet{chen2022week} applied an explainable artificial intelligence framework to predict the students' early learning performance and analyze their behavior in a virtual learning environment, which could be critical to address high failure rates in virtual learning courses. \\

\section{Method}

Several methods have been employed to assess student engagement in the virtual classroom \citep{christopoulos2018increasing, raes2020learning, rajalingam2021peer, sole2022data}. Quantitative measures such as login frequency, time spent on tasks, number of completed assignments, and participation in online discussions can provide objective data. Qualitative methods like student surveys, interviews, and analysis of discussion board posts can offer insights into students' subjective experiences and motivations. For instance, a survey might ask students about their level of interest in the course content, their perception of the personalization, and their willingness to actively participate. Interviews can delve deeper into students' reasons for engagement or disengagement and their suggestions for improvement.

However, these methods may be oversimplified and not reliable enough to indicate the student’s real engagement in the virtual classroom. This paper aims to assess student engagement through the eye movement paradigm with the eye-tracking technique embedded in Virtual Reality (VR) headset for the virtual classroom. This can result in a more comprehensive and robust experiment and analysis of the student’s real engagement in the virtual classroom.

Eye-tracking is to measure the estimated point of the gaze and the movement of an eye. The most popular and non-invasive way of eye-tracking is implemented by digital video cameras, calculating the eye position and movement through eye images in real time. A series of commercial VR headsets are equipped with eye-tracking mechanics, for instance, PICO 4 PRO, VIVE PRO EYE, and HP REVERB G2 OMNICEPT EDITION, where eye-tracking technique has been used in medicine-related research and applications. With the objective data provided by eye-tracking, a quantitative assessment of student engagement can be conducted through eye movement. \\

\subsection{Overall Application Requirements}
This section outlines the key requirements that define the overall characteristics of the application:
(1) Runtime Performance: The application must deliver an immersive experience with consistently high frame rates, ensuring smooth scene transitions and minimal eye-tracking response latency. The runtime performance should remain stable regardless of rendering quality.
(2) Realism and Graphics Quality: The application should be able to provide photorealistic rendering effects, lighting, and shadow.
(3) Accuracy: The application should provide accurate eye-tracking estimation and calculation to ensure the reliability of the testing results.
(4) Operability: The application should provide a user-friendly interface and instruments during the testing.
(5) Compatibility: The application should be compatible with various VR headsets, ensuring broad accessibility. \\ \\ 

\subsection{Integrated Structure Design of Application}
The integrated structure design of application is illustrated in Figure 1.

\begin{figure}[tb]
  \centering
  \rotatebox{270}{\includegraphics[height=9.0 cm]{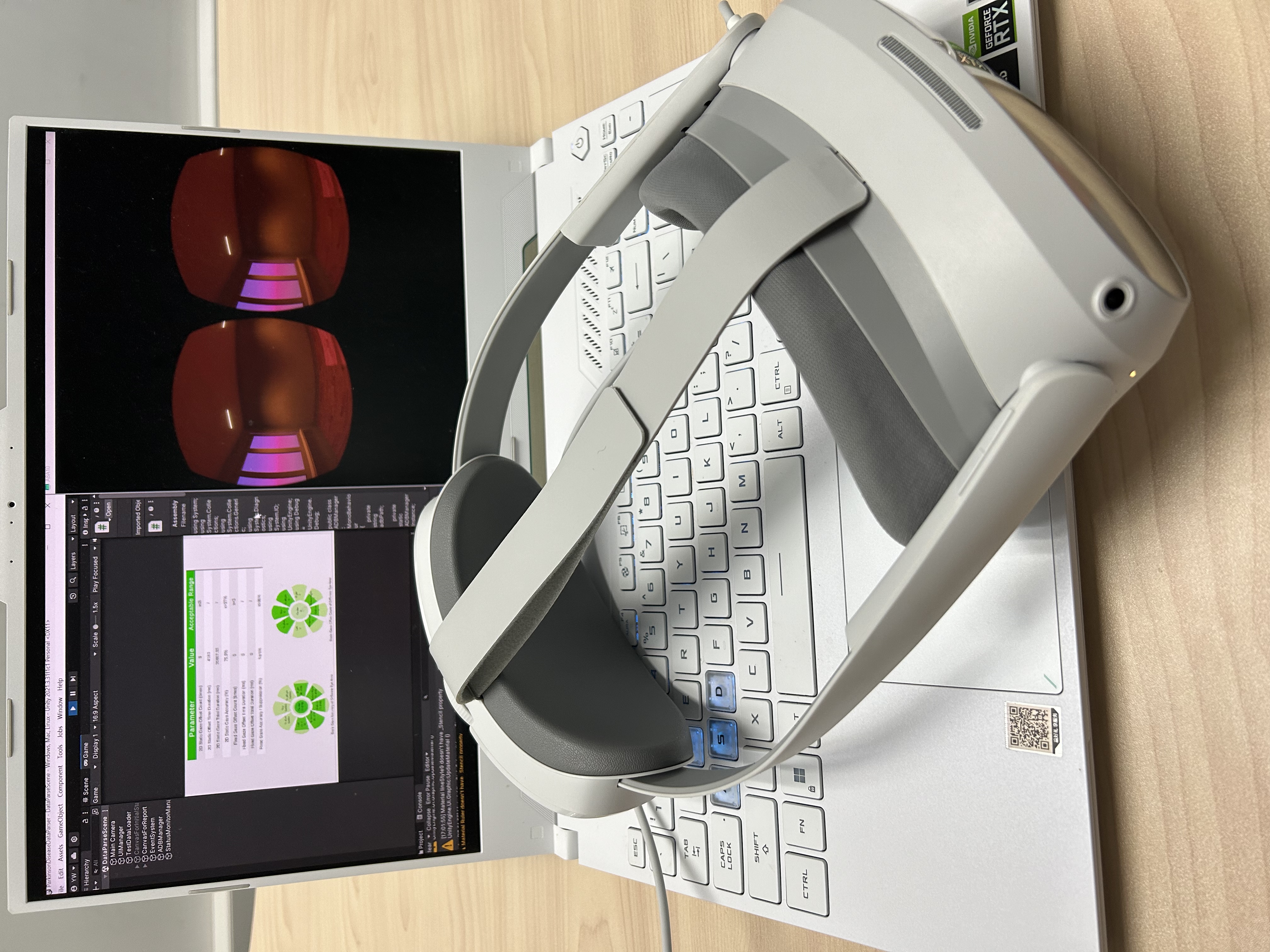}}
  \caption{Integrated structure design of application.}
  \label{fig:1}
\end{figure}

(1) Hardware Architecture:
This application requires a VR headset embedded with an eye-tracking device.
VR headset is used to render the contents in the virtual classroom and acquire the eye-tracking data. Then the report on student engagement in virtual classroom is automatically generated to provide analysis details.  

In our study, PICO VR 4 PRO is selected as the target platform of the application. A Windows PC is used for data analysis and result generation.

(2) Software Architecture:
This application is developed based on Unity’s Universal Render Pipeline. The overall Software Architecture Diagram is shown in Figure 2, which includes the following function modules:

• Eye-Tracking Testing Module: This module conducts eye-tracking tasks with user-friendly and concise guidance. During test, this module constantly
saves eye movement data, and eventually transfers the holistic data to Data Analysis Module.

• Data Capture Module: This module is responsible for capturing all various kinds of data of students in the virtual classroom during the testing procedure. Eye movement data includes: eye movement speed and gaze position on each frame, saccadic reaction time cost for each event, eye gaze offset count and accuracy and etc.

• Data Analysis \& Review Module: This module analyzes students’ eye movement according to the pre-defined criteria. The testing data and corresponding analysis report results will be saved as local files.

\begin{figure}[tb]
  \centering
  \includegraphics[height=7.0 cm]{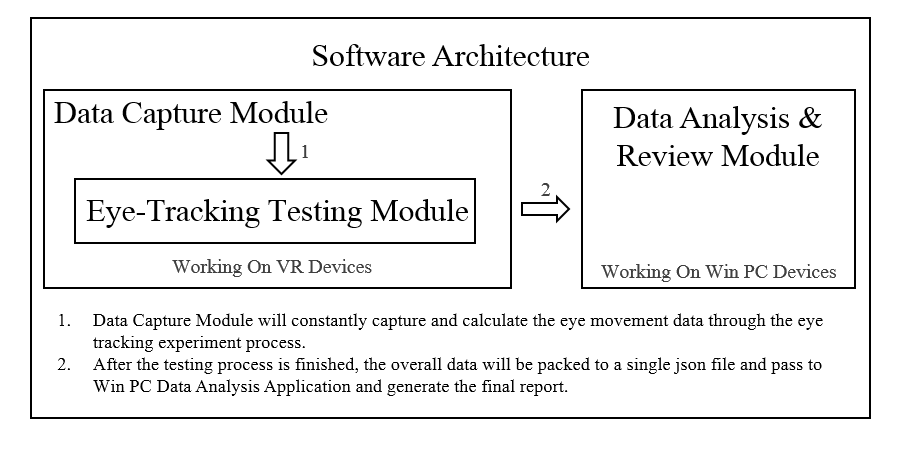}
  \caption{Overall Software Architecture Diagram.}
  \label{fig:2}
\end{figure}

\subsection{VR Eye-Tracking Application}

\subsubsection{Implementation of Eye-Tracking System}

(1) Eye-Tracking Data Provided by Pico VR SDK
PICO Unity Integration SDK provides a series of data related to user’s eye movement and head movement as follows:

• Human Head Position and Rotation: This data is the spatial position and the rotation of the VR headset, as well as the student’s head. It represents the current student’s head position and rotation status in Unity as a Matrix 4x4 format.

• Combined Eye Gazed Ray Vector: This data simulates the student’s gaze direction as a ray, while the student’s eye position is the starting point of the ray.

• Eye Openness Status: This data indicates whether the eyes of the student are currently open or closed, respectively.

(2) Other Data Calculated by Application

• World Space Eye Direction Ray Vector: This data represents the ray of the user’s eye gaze direction in the world reference axis as a ray Vector3. It is calculated through:

\begin{equation}
\begin{split}
\text{Vector3}[\text{WorldSpaceEyeGazeRay}] = \\
\text{Vector3}[\text{EyePos}] + \\
\text{Vector3}[\text{LocalSpaceEyeGazeRay}]
\end{split}
\end{equation}

• Screen Space Eye Gaze Position: This data indicates the coordinates of the
student’s current gaze point on screen space. It is calculated by the projection of the student’s gaze ray onto the VR screen space.

• Eye Direction Change Angle: This data is used to record the rotational change of the student’s gaze ray direction in each frame (14ms).

• Eye Angular Speed: This data represents the eye movement speed of the student in each frame, with units of °/s. It is calculated by:

\begin{equation}
EyeMovementSpeed (\frac{}{s}) = \frac{\Delta \theta}{\Delta t}
\end{equation}

Where $\Delta \theta$ denotes the Eye Movement Direction Angle Change in each fixed frame, and $\Delta t$ denotes the Time Duration during each fixed frame (14ms)

• Screen Space Eye Gaze Relative Area: This data is used to provide a summary description of the student’s current gaze position, including:

– Middle Area: Left, Right, Center

– Top Area: TopLeft, TopRight, Top

– Bottom Area: BottomLeft, BottomRight, Bottom

This is used to assist the program to analyze the user’s eye movement performance within different eye areas.

\subsubsection{Implementation of Testing Scenes}

(1) 2D Static Gaze Testing: 
In order to detect whether the student is gazing at a specific dot, this application utilizes the Unity Collision System. It first sends out a sphere-shaped ray cast through the student’s eye position along the eye gaze direction. Then, the collider of the red point collides with the eye ray cast, and the onEyeGazedAt() function will be triggered. Thus, a series of calculations will be carried out. As shown in Figure 3, the green line surrounding the red point illustrates its area of collision sphere.

\begin{figure}[tb]
  \centering
  \includegraphics[height=7.0 cm]{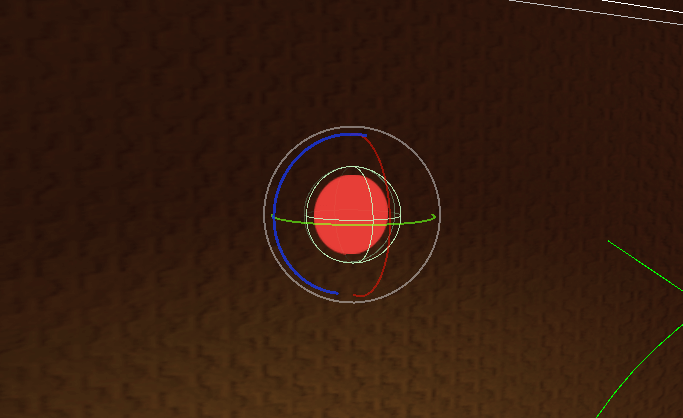}
  \caption{2D Red Point’s Collider.}
  \label{fig:3}
\end{figure}

The following data are collected during this testing process:

• 2D Calibration Gaze Offset Count

• 2D Calibration Gaze Offset Time Duration

• 2D Calibration Gaze Total Time Duration

• Eye Area Based Static Gaze Offset Count

• Eye Area Based Static Gaze Offset Time Duration

• Eye Area Based Static Total Time Duration

• Eye Area Based Saccadic Reaction Time Cost

• Eye Stimulate Event Data Point List

(2) 3D Dynamic Gaze Testing:

A Unity Plugin PathSystem is used in this application. It well supports straight line and bezier curves, thus a perfect choice of the implementation. Figure 4 displays the 3D dynamic gaze test path preview in Unity.

\begin{figure}[tb]
  \centering
  \includegraphics[height=6.0 cm]{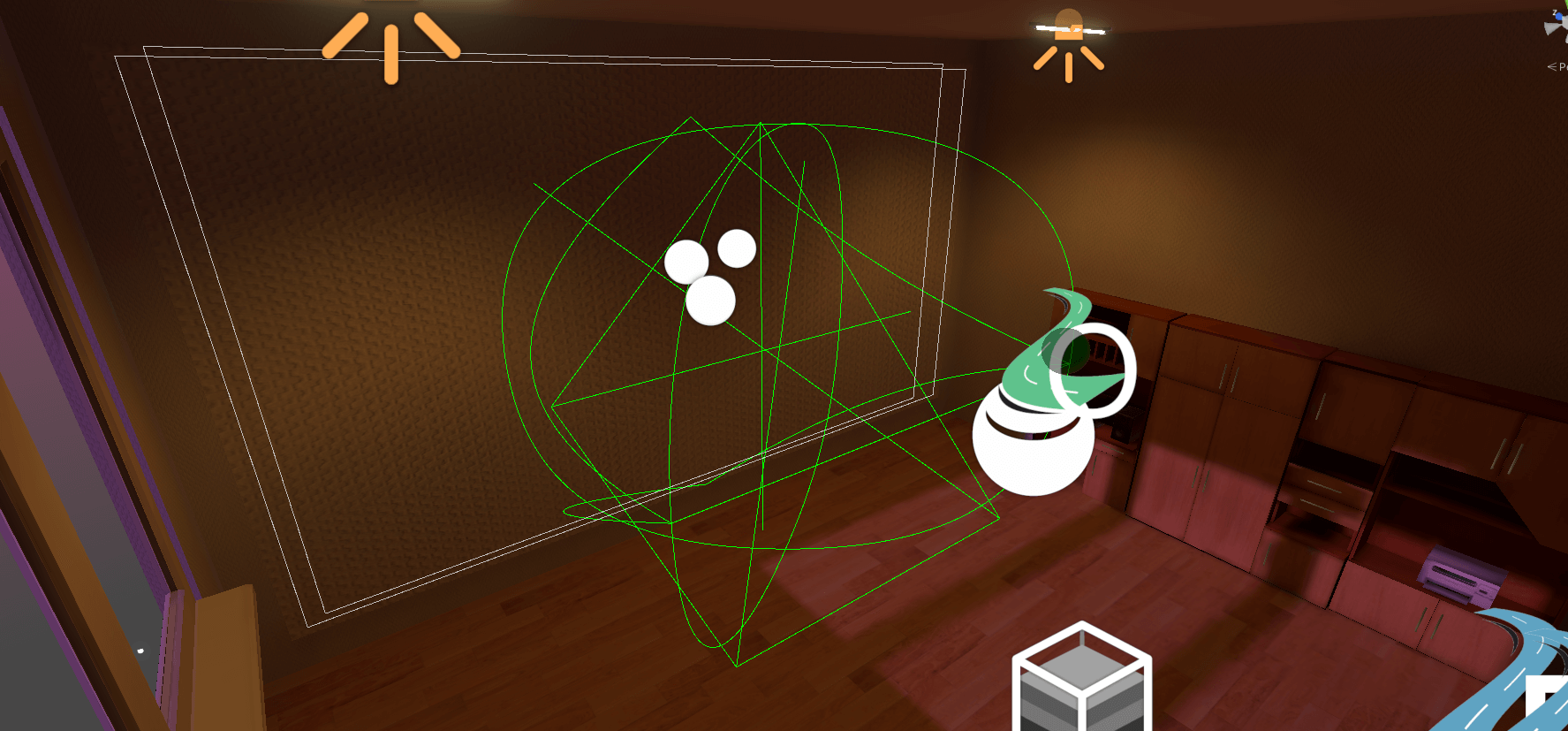}
  \caption{3D Dynamic Gaze Test Path Preview In Unity.}
  \label{fig:4}
\end{figure}

The following data are collected during this testing process:

• 3D Dynamic Gaze Offset Count

• 3D Dynamic Gaze Offset Time Duration

• 3D Dynamic Gaze Total Time Duration

• Eye Area Based 3D Dynamic Gaze Offset Count

• Eye Area Based 3D Dynamic Gaze Offset Time Duration

• Eye Area Based 3D Dynamic Total Time Duration

• Eye Area Based Saccadic Reaction Time Cost

• Eye Stimulate Event Data Point List

\subsection{Experimental Procedure and Research Participants}

In this section, we show the details of the experimental procedure for the case study. More specifically, we recruited eight participants for a case study on the effects of personalized learning in a virtual classroom. The study aimed to assess the impact of personalized learning on student performance. A questionnaire used to assess participants’ experiences and perceptions in the virtual classroom consisted of 11 questions, each measured on a five-point Likert scale (1 = ‘Strongly disagree’ to 5 = ‘Strongly agree’). These questions were designed to evaluate the participants' cognitive engagement, behavioral engagement, and affective engagement within the virtual learning environment. 

\subsubsection{Measurement}
All participants were asked to complete the same questionnaire, which was estimated to take approximately 5 minutes. The questionnaire first ask participants to report their background including age, gender, and education. Then, in order to analyze the participants' learning performance more comprehensively, we include 11 questions in the questionnaire to examine the following three dimensions: cognitive engagement, behavioral engagement, and affective engagement. Cognitive engagement refers to the extent to which student engages in thinking and using learning strategies in the virtual classroom \citep{fredricks2004school}. Behavioral engagement focuses on student's behavior during learning activities, such as active participation in discussions \citep{finn2012student}. Affective engagement refers to student's emotional response to the learning activity \citep{reeve2012self}.

\subsubsection{Questionnaire Questions}
The details of the 11 questions for cognitive engagement, behavioral engagement, and affective engagement in the questionnaire are shown below. Questions 1 to 3 are related the cognitive engagement, questions 4 to 6 are related to behavioral engagement, and question 7 to 11 are related to affective engagement.

\begin{enumerate}
    \item How well do you understand complex concepts when using a virtual classroom environment?
    \item Did the immersive virtual classroom experience help you understand the learning material?
    \item To what extent do you think virtual learning environments can enhance your thinking skills?
    \item Do you frequently and actively participate in discussions and interactions with the instructor in the virtual classroom?
    \item Are you more motivated to engage with learning content in a virtual classroom than in a traditional classroom?
    \item Are you able to stay focused for long periods of time during virtual classroom learning?
    \item How enjoyable do you find the learning experience in a virtual classroom?
    \item Has using a virtual classroom increased your enjoyment of classroom learning?
    \item How satisfied are you with the ability of the virtual classroom to meet your personal learning needs?
    \item Did you find it easy to operate and use the features of the virtual classroom?
    \item Do you think the technology in the virtual classroom is reliable and efficient?
\end{enumerate}

\section{Results and Discussion}
\subsection{Case Study Experimental Results}

\subsubsection{Questionnaire Results}
To analyze the questionnaire results, we conduct a comprehensive analysis of the responses. Questions 1 to 3 are categorized under cognitive engagement, questions 4 to 6 under behavioral engagement, and questions 7 to 11 under affective engagement. For each dimension, we calculate the mean value for each participant, which allows us to compare the relative levels of engagement across the three categories. The summarized results, presented in Table 1, provide an overview of the mean values for each type of engagement.

\begin{table}[h!]
\caption{Mean values for cognitive, behavioral, and affective engagement for eight participants.}
\centering
\begin{tabular}{cccc}
\toprule
\textbf{Participant} & \textbf{Cognitive engagement} & \textbf{Behavioral engagement} & \textbf{affective engagement} \\
\midrule
1 & 4.00 & 4.00 & 4.00 \\
2 & 5.00 & 5.00 & 5.00 \\
3 & 2.33 & 2.33 & 3.80 \\
4 & 4.00 & 4.33 & 3.80 \\
5 & 2.33 & 3.00 & 2.40 \\
6 & 5.00 & 4.00 & 4.60 \\
7 & 4.33 & 4.67 & 4.40 \\
8 & 4.00 & 3.00 & 4.00 \\
\bottomrule
\end{tabular}

\end{table}

The case study conducted on eight participants in the virtual environment reveals that those who received personalized learning experiences demonstrates higher levels of engagement as measured by increased participation in discussion. Students reported feeling that the personalized approach made the learning more relevant and interesting. 

More specifically, as shown in Table 1, individual mean scores for personalized learning via VR applications vary across cognitive, behavioral, and affective dimensions, highlighting the diversity in how participants experience personalized learning within virtual environments. For instance, Participant 2 demonstrated consistently high levels of engagement across all three categories, with mean values of 5.00, indicating a positive response to personalized learning approaches in VR. In contrast, Participant 5 showed comparatively lower levels, especially in the affective dimension.

The overall performance results indicate that most participants found personalized learning via VR applications beneficial, particularly in enhancing their engagement across cognitive, behavioral, and affective dimensions. The majority of participants scored consistently high in at least one or more engagement categories, suggesting that personalized learning in a VR environment was supportive for their learning needs.

\subsubsection{Automated Report Generation}
In this paper, we also develop and perform experimental data analysis and report generation application for assessing student engagement in the virtual classroom. This application reads experimental data result files in JSON format and presents the data results as tables and various types of charts. The general information module (Figure 5) is used to show the basic information of users and tests, including name, age, gender, and the overall test score. Static gaze test data (Figure 6) describes the performance of 2D Static Gaze and Disturbance-based Gaze. The two charts at the bottom represent the static gaze accuracy and offset count, comparing different eye areas. Dynamic gaze test data describes the performance of 3D Dynamic Gaze (Figure 7, Figure 8). The line charts in below graphs are used to represent the fitting degree between the user’s gaze movement path and the target object’s movement path. The two charts at the bottom represent the dyanmic gaze accuracy and offset count corresponding to different eye areas. Saccadic test data describes the user’s reaction time cost/speed during saccadic movements (Figure 9, Figure 10). In the line chart below, the straight green line represents the user’s current eye movement speed. The red points represent eye stimulation events. This chart is used to reflect the impact of visual stimulation events on the user’s eye movement. Figure 11 gives overall evaluation to user’s eye movement performance.

\begin{figure}[tb]
  \centering
  \includegraphics[height=3.5 cm]{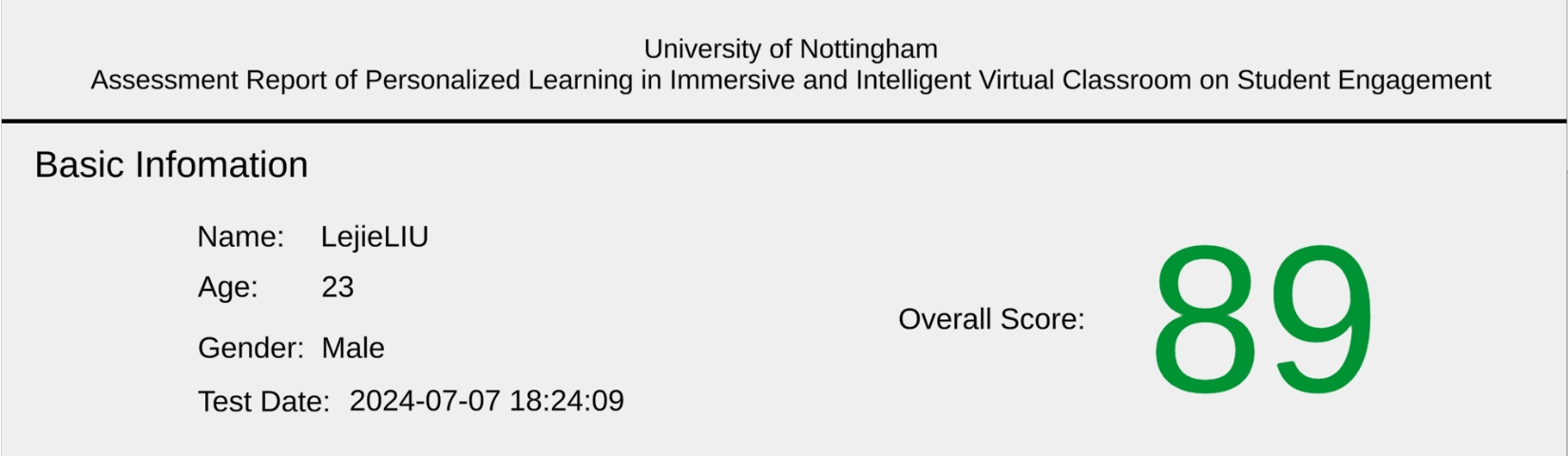}
  \caption{Report of General Information.}
  \label{fig:5}
\end{figure}

\begin{figure}[tb]
  \centering
  \includegraphics[height=9.0 cm]{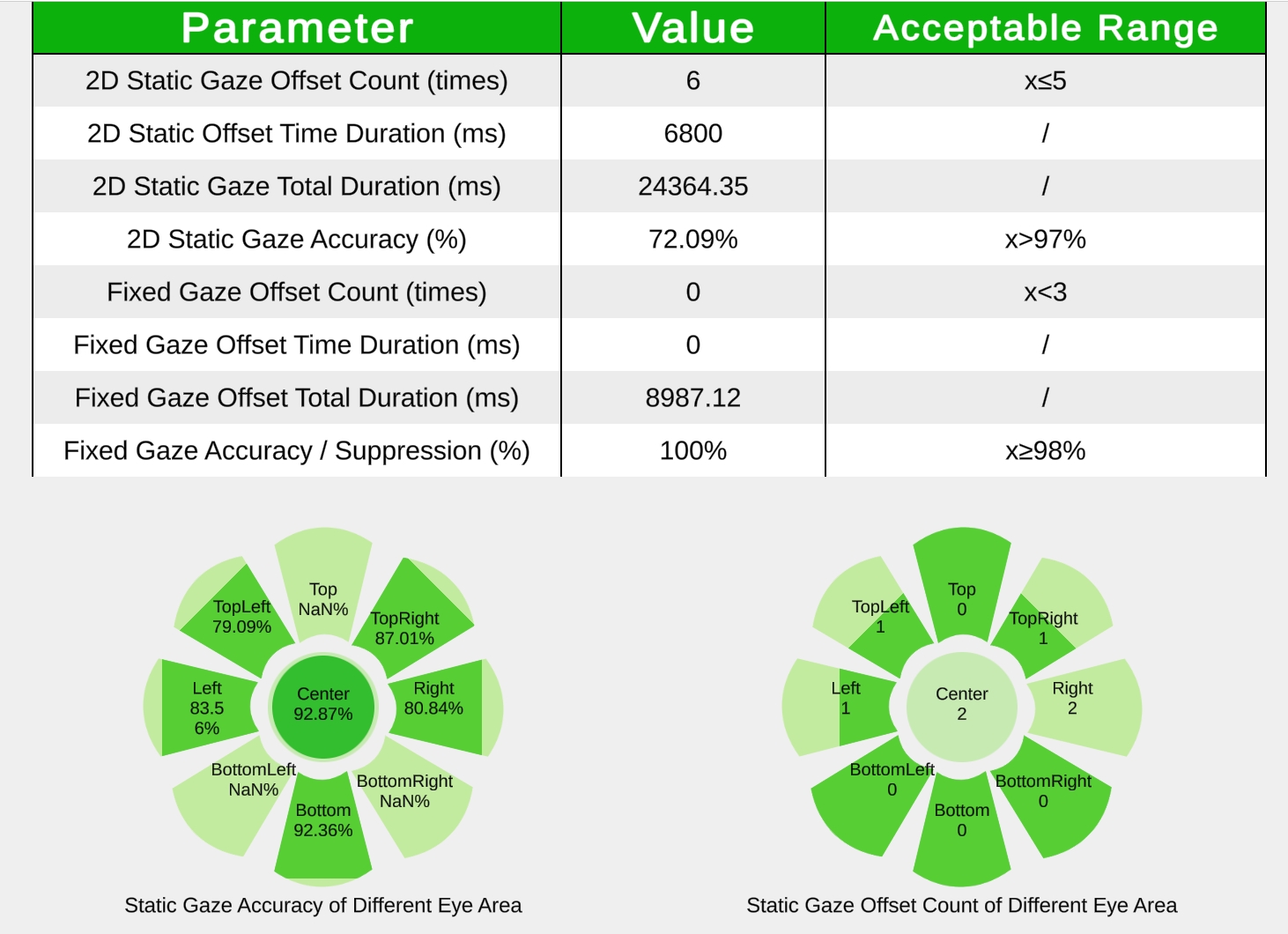}
  \caption{Report of Static Gaze Data.}
  \label{fig:6}
\end{figure}

\begin{figure}[tb]
  \centering
  \includegraphics[height=6.5 cm]{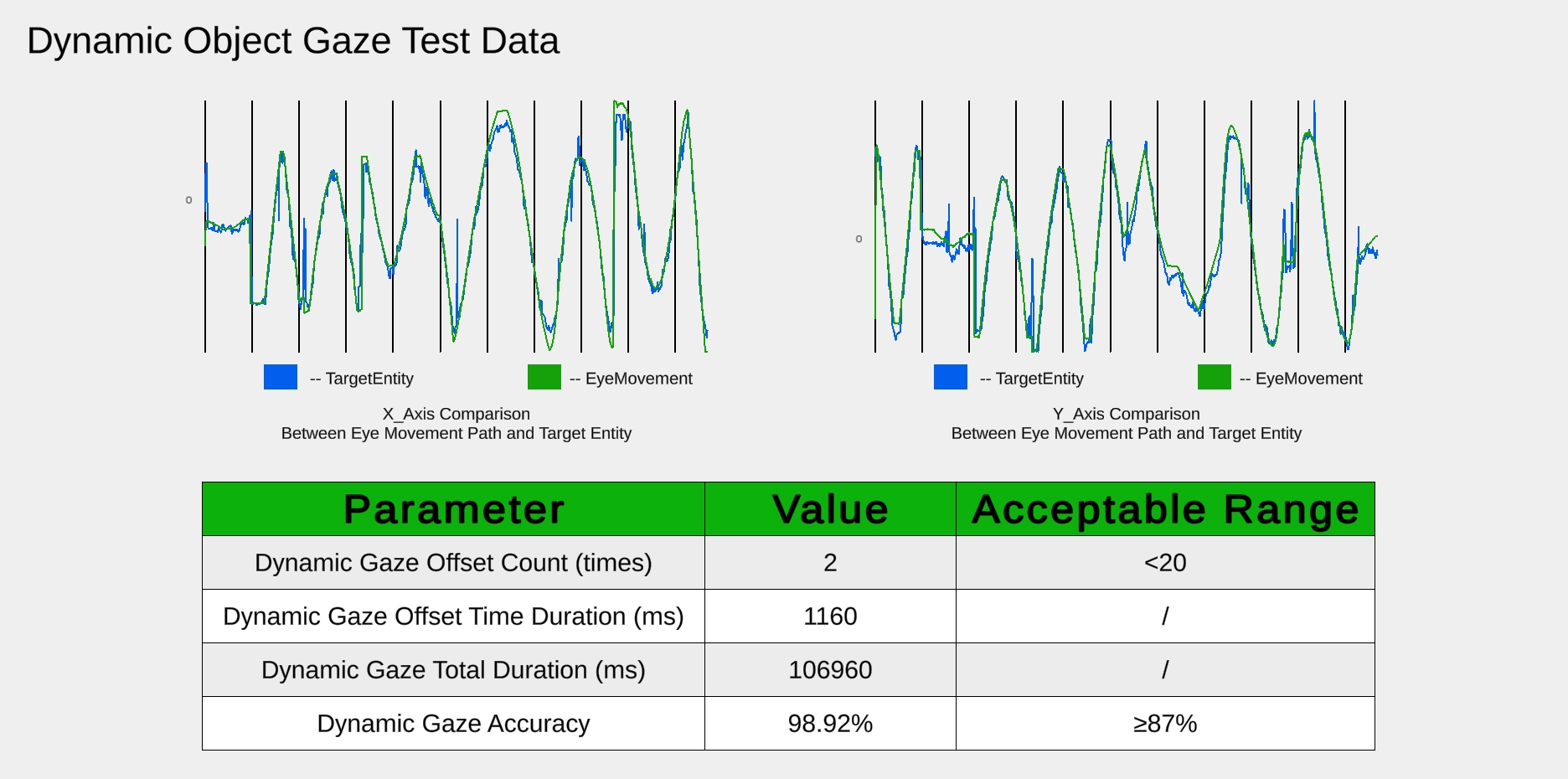}
  \caption{Report of Dynamic Gaze Test Data 1.}
  \label{fig:7}
\end{figure}

\begin{figure}[tb]
  \centering
  \includegraphics[height=4.1 cm]{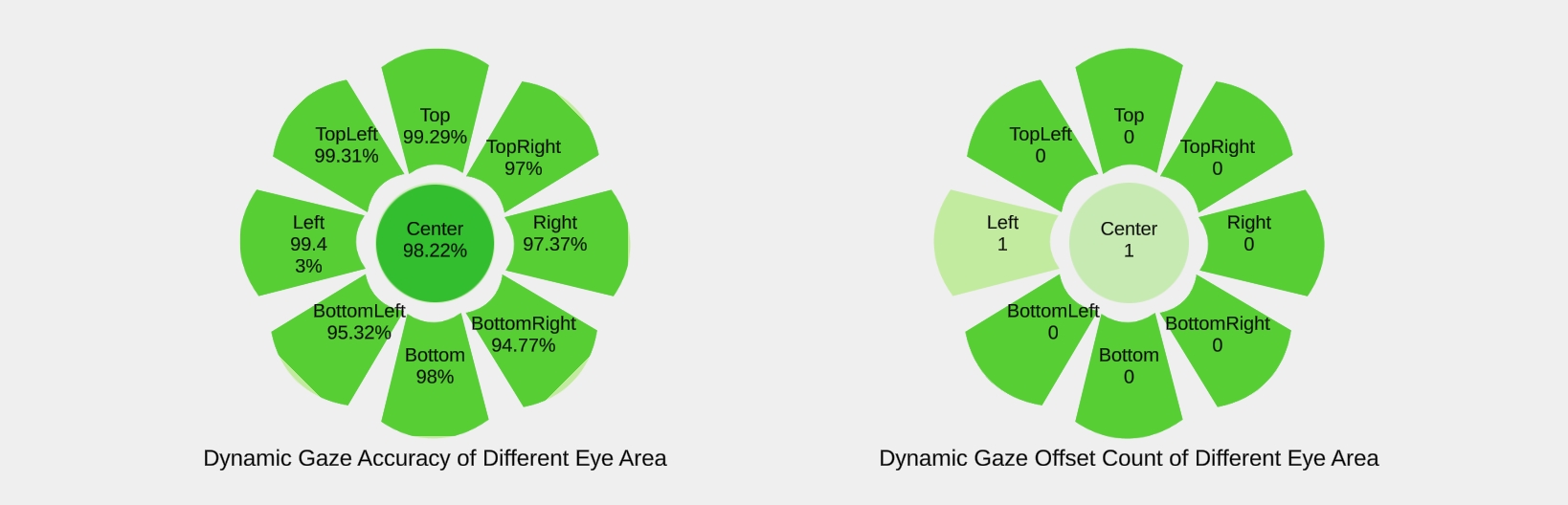}
  \caption{Report of Dynamic Gaze Test Data 2.}
  \label{fig:8-1}
\end{figure}

\begin{figure}[tb]
  \centering
  \includegraphics[height=5.0 cm]{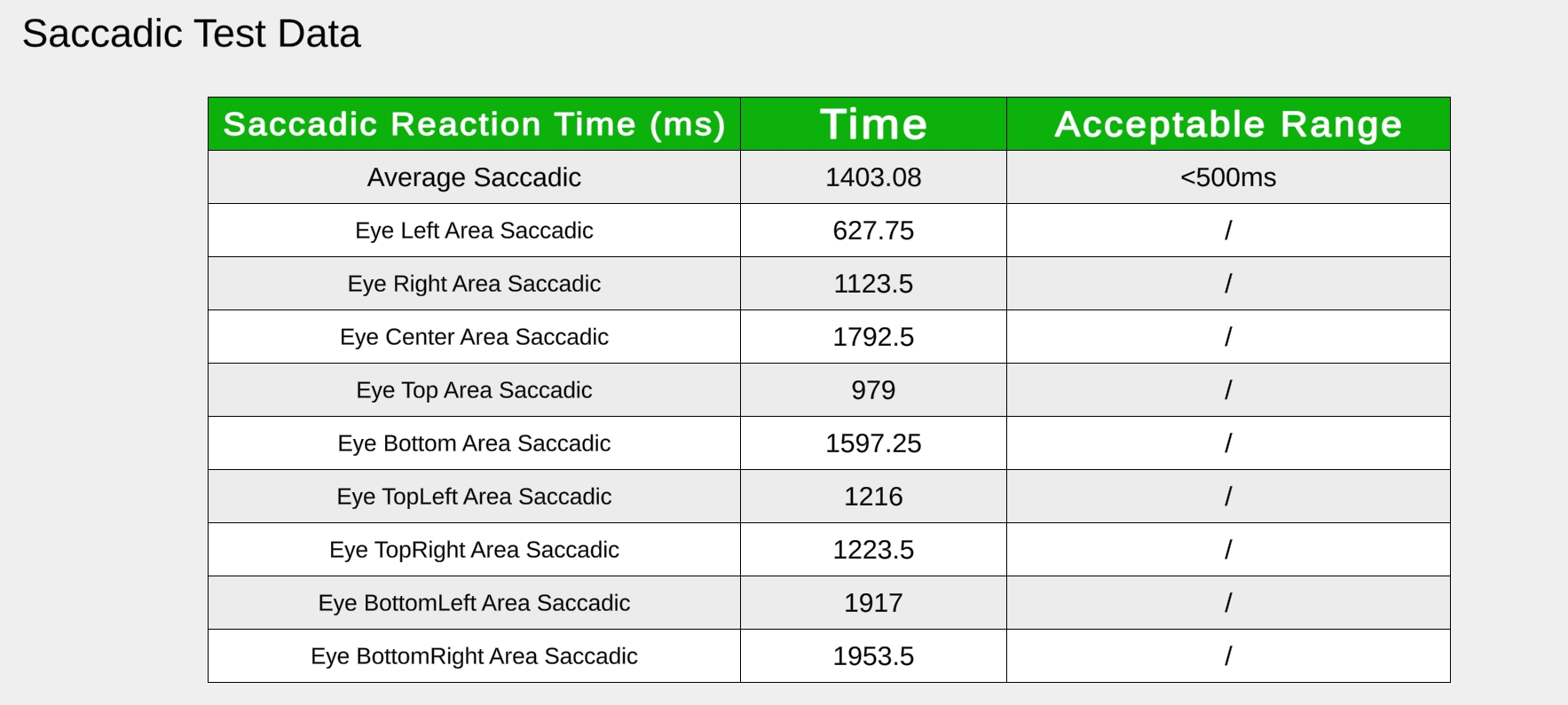}
  \caption{Report of Saccadic Test Data 1.}
  \label{fig:8-2}
\end{figure}

\begin{figure}[tb]
  \centering
  \includegraphics[height=4.99 cm]{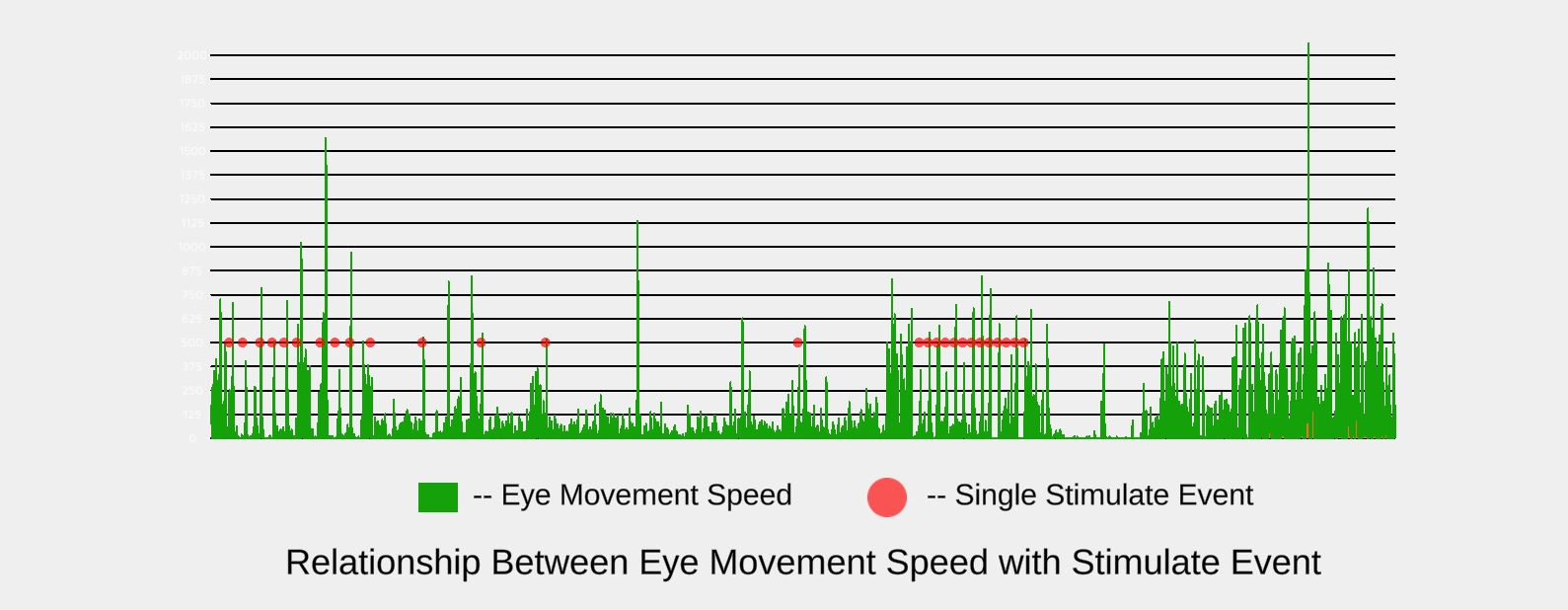}
  \caption{Report of Saccadic Test Data 2.}
  \label{fig:8-3}
\end{figure}

\begin{figure}[tb]
  \centering
  \includegraphics[height=4.8 cm]{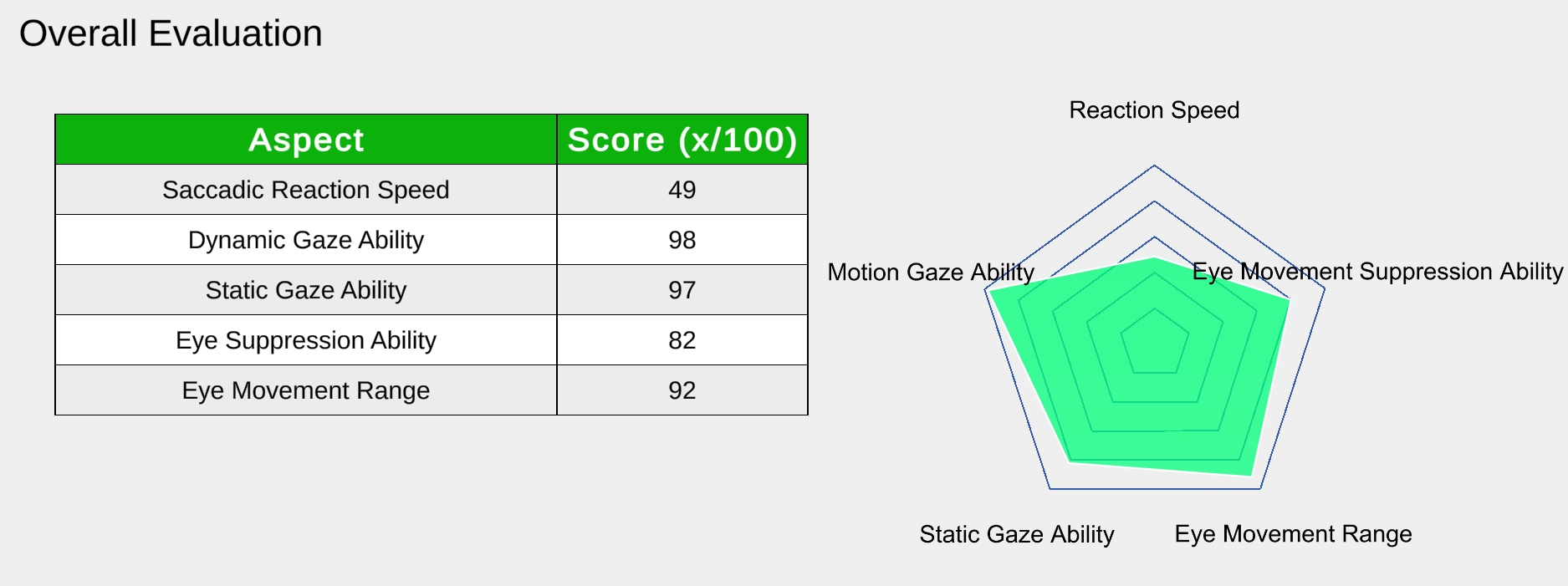}
  \caption{Report of Overall Evaluation.}
  \label{fig:8-4}
\end{figure}

\subsection{Findings and Analysis}

In response to RQ1, the case study demonstrates that 3D eye-tracking paradigms can effectively assess personalized learning in virtual classrooms, particularly in measuring student engagement. Eye-tracking data, such as gaze patterns and fixation duration, provides valuable insights into how students interacted with learning materials and their levels of attentiveness. For RQ2, the analysis of eye-tracking data reveals several key metrics that are particularly useful for assessing student engagement in virtual classrooms. Among these, saccadic reaction time which is defined as the time taken for the eyes to jump from one focus point to another is emerged as a critical indicator of cognitive processing and attention.

Based on the findings, several implications and recommendations can be drawn for educators and instructional designers. Firstly, it is crucial to invest in technologies and platforms that support personalized learning capabilities. Secondly, teachers need training and resources to effectively implement personalized learning strategies and monitor student engagement.

Educators should also communicate the benefits and expectations of personalized learning to students to enhance their acceptance and active participation. Ongoing assessment and feedback loops should be established to continuously refine the personalized learning experiences based on student responses.

For example, providing regular progress reports to students and allowing them to have input in shaping their learning paths can increase their sense of ownership and engagement.  \\

\subsection{Limitations and Future Research Directions}
Despite the promising findings, the limitation of the research should also be acknowledged. The relatively small sample size used in the case study may limit the generalizability of the results. To enhance the applicability of these findings across different educational contexts, future research should aim to replicate the study with larger and more diverse student populations. The assessment of personalized learning in intelligent virtual classrooms, particularly in terms of its impact on student engagement, is a complex but essential area of research. While this study has demonstrated the potential of eye-tracking technology for evaluating engagement, further exploration is needed to understand the long-term effects of personalized learning. Future studies should also examine diverse subject areas, varying educational levels, and different age groups to gain a broader understanding of how personalized learning affects engagement and educational outcomes. By continuously refining personalized learning approaches and expanding research across different contexts, we can create more dynamic and effective virtual learning environments. This ongoing evaluation will help meet the unique needs of individual students, ultimately fostering deeper engagement and better academic performance. \\

\section{Conclusion}
The assessment of personalized learning in the intelligent virtual classroom on student engagement is a complex but essential area of research. This study has demonstrated that careful implementation and ongoing evaluation are necessary to ensure its effectiveness. Future research should explore long-term effects, diverse subject areas, and different age groups to further validate and expand the understanding of this relationship. By continuously improving personalized learning approaches and assessing their impact on student engagement, we can create more dynamic and effective virtual learning environments that meet the unique needs of each student and foster better educational outcomes.


\subsection*{Data Availability}
The data that support the findings of this study are available from the corresponding author, upon reasonable request.

\subsection*{Conflict of interest}
There is no relevant financial or non-financial competing interests to report.

\bibliographystyle{apacite}


\begin{thebibliography}{}

\bibitem [\protect \citeauthoryear {%
Alamri%
, Lowell%
, Watson%
\BCBL {}\ \BBA {} Watson%
}{%
Alamri%
\ \protect \BOthers {.}}{%
{\protect \APACyear {2020}}%
}]{%
alamri2020using}
\APACinsertmetastar {%
alamri2020using}%
\begin{APACrefauthors}%
Alamri, H.%
, Lowell, V.%
, Watson, W.%
\BCBL {}\ \BBA {} Watson, S\BPBI L.%
\end{APACrefauthors}%
\unskip\
\newblock
\APACrefYearMonthDay{2020}{}{}.
\newblock
{\BBOQ}\APACrefatitle {Using personalized learning as an instructional approach to motivate learners in online higher education: Learner self-determination and intrinsic motivation} {Using personalized learning as an instructional approach to motivate learners in online higher education: Learner self-determination and intrinsic motivation}.{\BBCQ}
\newblock
\APACjournalVolNumPages{Journal of Research on Technology in Education}{52}{3}{322--352}.
\PrintBackRefs{\CurrentBib}

\bibitem [\protect \citeauthoryear {%
Anderson%
\ \BBA {} Dron%
}{%
Anderson%
\ \BBA {} Dron%
}{%
{\protect \APACyear {2011}}%
}]{%
anderson2011three}
\APACinsertmetastar {%
anderson2011three}%
\begin{APACrefauthors}%
Anderson, T.%
\BCBT {}\ \BBA {} Dron, J.%
\end{APACrefauthors}%
\unskip\
\newblock
\APACrefYearMonthDay{2011}{}{}.
\newblock
{\BBOQ}\APACrefatitle {Three generations of distance education pedagogy} {Three generations of distance education pedagogy}.{\BBCQ}
\newblock
\APACjournalVolNumPages{International Review of Research in Open and Distributed Learning}{12}{3}{80--97}.
\PrintBackRefs{\CurrentBib}

\bibitem [\protect \citeauthoryear {%
Basham%
, Hall%
, Carter~Jr%
\BCBL {}\ \BBA {} Stahl%
}{%
Basham%
\ \protect \BOthers {.}}{%
{\protect \APACyear {2016}}%
}]{%
basham2016operationalized}
\APACinsertmetastar {%
basham2016operationalized}%
\begin{APACrefauthors}%
Basham, J\BPBI D.%
, Hall, T\BPBI E.%
, Carter~Jr, R\BPBI A.%
\BCBL {}\ \BBA {} Stahl, W\BPBI M.%
\end{APACrefauthors}%
\unskip\
\newblock
\APACrefYearMonthDay{2016}{}{}.
\newblock
{\BBOQ}\APACrefatitle {An operationalized understanding of personalized learning} {An operationalized understanding of personalized learning}.{\BBCQ}
\newblock
\APACjournalVolNumPages{Journal of Special Education Technology}{31}{3}{126--136}.
\PrintBackRefs{\CurrentBib}

\bibitem [\protect \citeauthoryear {%
H\BHBI C.~Chen%
\ \protect \BOthers {.}}{%
H\BHBI C.~Chen%
\ \protect \BOthers {.}}{%
{\protect \APACyear {2022}}%
}]{%
chen2022week}
\APACinsertmetastar {%
chen2022week}%
\begin{APACrefauthors}%
Chen, H\BHBI C.%
, Prasetyo, E.%
, Tseng, S\BHBI S.%
, Putra, K\BPBI T.%
, Kusumawardani, S\BPBI S.%
\BCBL {}\ \BBA {} Weng, C\BHBI E.%
\end{APACrefauthors}%
\unskip\
\newblock
\APACrefYearMonthDay{2022}{}{}.
\newblock
{\BBOQ}\APACrefatitle {Week-wise student performance early prediction in virtual learning environment using a deep explainable artificial intelligence} {Week-wise student performance early prediction in virtual learning environment using a deep explainable artificial intelligence}.{\BBCQ}
\newblock
\APACjournalVolNumPages{Applied Sciences}{12}{4}{1885}.
\PrintBackRefs{\CurrentBib}

\bibitem [\protect \citeauthoryear {%
L.~Chen%
, Chen%
\BCBL {}\ \BBA {} Lin%
}{%
L.~Chen%
\ \protect \BOthers {.}}{%
{\protect \APACyear {2020}}%
}]{%
chen2020artificial}
\APACinsertmetastar {%
chen2020artificial}%
\begin{APACrefauthors}%
Chen, L.%
, Chen, P.%
\BCBL {}\ \BBA {} Lin, Z.%
\end{APACrefauthors}%
\unskip\
\newblock
\APACrefYearMonthDay{2020}{}{}.
\newblock
{\BBOQ}\APACrefatitle {Artificial intelligence in education: A review} {Artificial intelligence in education: A review}.{\BBCQ}
\newblock
\APACjournalVolNumPages{Ieee Access}{8}{}{75264--75278}.
\PrintBackRefs{\CurrentBib}

\bibitem [\protect \citeauthoryear {%
Christopoulos%
, Conrad%
\BCBL {}\ \BBA {} Shukla%
}{%
Christopoulos%
\ \protect \BOthers {.}}{%
{\protect \APACyear {2018}}%
}]{%
christopoulos2018increasing}
\APACinsertmetastar {%
christopoulos2018increasing}%
\begin{APACrefauthors}%
Christopoulos, A.%
, Conrad, M.%
\BCBL {}\ \BBA {} Shukla, M.%
\end{APACrefauthors}%
\unskip\
\newblock
\APACrefYearMonthDay{2018}{}{}.
\newblock
{\BBOQ}\APACrefatitle {Increasing student engagement through virtual interactions: How?} {Increasing student engagement through virtual interactions: How?}{\BBCQ}
\newblock
\APACjournalVolNumPages{Virtual Reality}{22}{4}{353--369}.
\PrintBackRefs{\CurrentBib}

\bibitem [\protect \citeauthoryear {%
Demszky%
, Liu%
, Hill%
, Sanghi%
\BCBL {}\ \BBA {} Chung%
}{%
Demszky%
\ \protect \BOthers {.}}{%
{\protect \APACyear {2024}}%
}]{%
demszky2024automated}
\APACinsertmetastar {%
demszky2024automated}%
\begin{APACrefauthors}%
Demszky, D.%
, Liu, J.%
, Hill, H\BPBI C.%
, Sanghi, S.%
\BCBL {}\ \BBA {} Chung, A.%
\end{APACrefauthors}%
\unskip\
\newblock
\APACrefYearMonthDay{2024}{}{}.
\newblock
{\BBOQ}\APACrefatitle {Automated Feedback Improves Teachers’ Questioning Quality in Brick-and-Mortar Classrooms: Opportunities for Further Enhancement} {Automated feedback improves teachers’ questioning quality in brick-and-mortar classrooms: Opportunities for further enhancement}.{\BBCQ}
\newblock
\APACjournalVolNumPages{Computers \& Education}{}{}{105183}.
\PrintBackRefs{\CurrentBib}

\bibitem [\protect \citeauthoryear {%
Ennouamani%
\ \BBA {} Mahani%
}{%
Ennouamani%
\ \BBA {} Mahani%
}{%
{\protect \APACyear {2017}}%
}]{%
ennouamani2017overview}
\APACinsertmetastar {%
ennouamani2017overview}%
\begin{APACrefauthors}%
Ennouamani, S.%
\BCBT {}\ \BBA {} Mahani, Z.%
\end{APACrefauthors}%
\unskip\
\newblock
\APACrefYearMonthDay{2017}{}{}.
\newblock
{\BBOQ}\APACrefatitle {An overview of adaptive e-learning systems} {An overview of adaptive e-learning systems}.{\BBCQ}
\newblock
\BIn{} \APACrefbtitle {2017 eighth international conference on intelligent computing and information systems (ICICIS)} {2017 eighth international conference on intelligent computing and information systems (icicis)}\ (\BPGS\ 342--347).
\PrintBackRefs{\CurrentBib}

\bibitem [\protect \citeauthoryear {%
Finn%
\ \BBA {} Zimmer%
}{%
Finn%
\ \BBA {} Zimmer%
}{%
{\protect \APACyear {2012}}%
}]{%
finn2012student}
\APACinsertmetastar {%
finn2012student}%
\begin{APACrefauthors}%
Finn, J\BPBI D.%
\BCBT {}\ \BBA {} Zimmer, K\BPBI S.%
\end{APACrefauthors}%
\unskip\
\newblock
\APACrefYearMonthDay{2012}{}{}.
\newblock
{\BBOQ}\APACrefatitle {Student engagement: What is it? Why does it matter?} {Student engagement: What is it? why does it matter?}{\BBCQ}
\newblock
\BIn{} \APACrefbtitle {Handbook of research on student engagement} {Handbook of research on student engagement}\ (\BPGS\ 97--131).
\newblock
\APACaddressPublisher{}{Springer}.
\PrintBackRefs{\CurrentBib}

\bibitem [\protect \citeauthoryear {%
Fredricks%
, Blumenfeld%
\BCBL {}\ \BBA {} Paris%
}{%
Fredricks%
\ \protect \BOthers {.}}{%
{\protect \APACyear {2004}}%
}]{%
fredricks2004school}
\APACinsertmetastar {%
fredricks2004school}%
\begin{APACrefauthors}%
Fredricks, J\BPBI A.%
, Blumenfeld, P\BPBI C.%
\BCBL {}\ \BBA {} Paris, A\BPBI H.%
\end{APACrefauthors}%
\unskip\
\newblock
\APACrefYearMonthDay{2004}{}{}.
\newblock
{\BBOQ}\APACrefatitle {School engagement: Potential of the concept, state of the evidence} {School engagement: Potential of the concept, state of the evidence}.{\BBCQ}
\newblock
\APACjournalVolNumPages{Review of educational research}{74}{1}{59--109}.
\PrintBackRefs{\CurrentBib}

\bibitem [\protect \citeauthoryear {%
A\BPBI Y.~Huang%
, Lu%
\BCBL {}\ \BBA {} Yang%
}{%
A\BPBI Y.~Huang%
\ \protect \BOthers {.}}{%
{\protect \APACyear {2023}}%
}]{%
huang2023effects}
\APACinsertmetastar {%
huang2023effects}%
\begin{APACrefauthors}%
Huang, A\BPBI Y.%
, Lu, O\BPBI H.%
\BCBL {}\ \BBA {} Yang, S\BPBI J.%
\end{APACrefauthors}%
\unskip\
\newblock
\APACrefYearMonthDay{2023}{}{}.
\newblock
{\BBOQ}\APACrefatitle {Effects of artificial Intelligence--Enabled personalized recommendations on learners’ learning engagement, motivation, and outcomes in a flipped classroom} {Effects of artificial intelligence--enabled personalized recommendations on learners’ learning engagement, motivation, and outcomes in a flipped classroom}.{\BBCQ}
\newblock
\APACjournalVolNumPages{Computers \& Education}{194}{}{104684}.
\PrintBackRefs{\CurrentBib}

\bibitem [\protect \citeauthoryear {%
R.~Huang%
\ \protect \BOthers {.}}{%
R.~Huang%
\ \protect \BOthers {.}}{%
{\protect \APACyear {2020}}%
}]{%
huang2020disrupted}
\APACinsertmetastar {%
huang2020disrupted}%
\begin{APACrefauthors}%
Huang, R.%
, Tlili, A.%
, Chang, T\BHBI W.%
, Zhang, X.%
, Nascimbeni, F.%
\BCBL {}\ \BBA {} Burgos, D.%
\end{APACrefauthors}%
\unskip\
\newblock
\APACrefYearMonthDay{2020}{}{}.
\newblock
{\BBOQ}\APACrefatitle {Disrupted classes, undisrupted learning during COVID-19 outbreak in China: application of open educational practices and resources} {Disrupted classes, undisrupted learning during covid-19 outbreak in china: application of open educational practices and resources}.{\BBCQ}
\newblock
\APACjournalVolNumPages{Smart Learning Environments}{7}{}{1--15}.
\PrintBackRefs{\CurrentBib}

\bibitem [\protect \citeauthoryear {%
Y.~Huang%
, Richter%
, Kleickmann%
, Scheiter%
\BCBL {}\ \BBA {} Richter%
}{%
Y.~Huang%
\ \protect \BOthers {.}}{%
{\protect \APACyear {2023}}%
}]{%
huang2023body}
\APACinsertmetastar {%
huang2023body}%
\begin{APACrefauthors}%
Huang, Y.%
, Richter, E.%
, Kleickmann, T.%
, Scheiter, K.%
\BCBL {}\ \BBA {} Richter, D.%
\end{APACrefauthors}%
\unskip\
\newblock
\APACrefYearMonthDay{2023}{}{}.
\newblock
{\BBOQ}\APACrefatitle {Body in motion, attention in focus: A virtual reality study on teachers' movement patterns and noticing} {Body in motion, attention in focus: A virtual reality study on teachers' movement patterns and noticing}.{\BBCQ}
\newblock
\APACjournalVolNumPages{Computers \& Education}{206}{}{104912}.
\PrintBackRefs{\CurrentBib}

\bibitem [\protect \citeauthoryear {%
Jongbloed%
, Chaker%
\BCBL {}\ \BBA {} Lavou{\'e}%
}{%
Jongbloed%
\ \protect \BOthers {.}}{%
{\protect \APACyear {2024}}%
}]{%
jongbloed2024immersive}
\APACinsertmetastar {%
jongbloed2024immersive}%
\begin{APACrefauthors}%
Jongbloed, J.%
, Chaker, R.%
\BCBL {}\ \BBA {} Lavou{\'e}, E.%
\end{APACrefauthors}%
\unskip\
\newblock
\APACrefYearMonthDay{2024}{}{}.
\newblock
{\BBOQ}\APACrefatitle {Immersive procedural training in virtual reality: A systematic literature review} {Immersive procedural training in virtual reality: A systematic literature review}.{\BBCQ}
\newblock
\APACjournalVolNumPages{Computers \& Education}{}{}{105124}.
\PrintBackRefs{\CurrentBib}

\bibitem [\protect \citeauthoryear {%
Katiyar%
\ \protect \BOthers {.}}{%
Katiyar%
\ \protect \BOthers {.}}{%
{\protect \APACyear {2024}}%
}]{%
katiyar2024ai}
\APACinsertmetastar {%
katiyar2024ai}%
\begin{APACrefauthors}%
Katiyar, N.%
, Awasthi, M\BPBI V\BPBI K.%
, Pratap, R.%
, Mishra, M\BPBI K.%
, Shukla, M\BPBI N.%
, Tiwari, M.%
\BCBL {}\ \BOthersPeriod {.}\end{APACrefauthors}%
\unskip\
\newblock
\APACrefYearMonthDay{2024}{}{}.
\newblock
{\BBOQ}\APACrefatitle {Ai-Driven Personalized Learning Systems: Enhancing Educational Effectiveness} {Ai-driven personalized learning systems: Enhancing educational effectiveness}.{\BBCQ}
\newblock
\APACjournalVolNumPages{Educational Administration: Theory and Practice}{30}{5}{11514--11524}.
\PrintBackRefs{\CurrentBib}

\bibitem [\protect \citeauthoryear {%
Kavitha%
\ \BBA {} Lohani%
}{%
Kavitha%
\ \BBA {} Lohani%
}{%
{\protect \APACyear {2019}}%
}]{%
kavitha2019critical}
\APACinsertmetastar {%
kavitha2019critical}%
\begin{APACrefauthors}%
Kavitha, V.%
\BCBT {}\ \BBA {} Lohani, R.%
\end{APACrefauthors}%
\unskip\
\newblock
\APACrefYearMonthDay{2019}{}{}.
\newblock
{\BBOQ}\APACrefatitle {A critical study on the use of artificial intelligence, e-Learning technology and tools to enhance the learners experience} {A critical study on the use of artificial intelligence, e-learning technology and tools to enhance the learners experience}.{\BBCQ}
\newblock
\APACjournalVolNumPages{Cluster Computing}{22}{Suppl 3}{6985--6989}.
\PrintBackRefs{\CurrentBib}

\bibitem [\protect \citeauthoryear {%
Kumar%
\ \protect \BOthers {.}}{%
Kumar%
\ \protect \BOthers {.}}{%
{\protect \APACyear {2023}}%
}]{%
kumar2023exploring}
\APACinsertmetastar {%
kumar2023exploring}%
\begin{APACrefauthors}%
Kumar, D.%
, Haque, A.%
, Mishra, K.%
, Islam, F.%
, Mishra, B\BPBI K.%
\BCBL {}\ \BBA {} Ahmad, S.%
\end{APACrefauthors}%
\unskip\
\newblock
\APACrefYearMonthDay{2023}{}{}.
\newblock
{\BBOQ}\APACrefatitle {Exploring the transformative role of artificial intelligence and metaverse in education: A comprehensive review} {Exploring the transformative role of artificial intelligence and metaverse in education: A comprehensive review}.{\BBCQ}
\newblock
\APACjournalVolNumPages{Metaverse Basic and Applied Research}{2}{}{55--55}.
\PrintBackRefs{\CurrentBib}

\bibitem [\protect \citeauthoryear {%
Lin%
, Yeh%
, Hung%
\BCBL {}\ \BBA {} Chang%
}{%
Lin%
\ \protect \BOthers {.}}{%
{\protect \APACyear {2013}}%
}]{%
lin2013data}
\APACinsertmetastar {%
lin2013data}%
\begin{APACrefauthors}%
Lin, C\BPBI F.%
, Yeh, Y\BHBI c.%
, Hung, Y\BPBI H.%
\BCBL {}\ \BBA {} Chang, R\BPBI I.%
\end{APACrefauthors}%
\unskip\
\newblock
\APACrefYearMonthDay{2013}{}{}.
\newblock
{\BBOQ}\APACrefatitle {Data mining for providing a personalized learning path in creativity: An application of decision trees} {Data mining for providing a personalized learning path in creativity: An application of decision trees}.{\BBCQ}
\newblock
\APACjournalVolNumPages{Computers \& Education}{68}{}{199--210}.
\PrintBackRefs{\CurrentBib}

\bibitem [\protect \citeauthoryear {%
Mousavinasab%
\ \protect \BOthers {.}}{%
Mousavinasab%
\ \protect \BOthers {.}}{%
{\protect \APACyear {2021}}%
}]{%
mousavinasab2021intelligent}
\APACinsertmetastar {%
mousavinasab2021intelligent}%
\begin{APACrefauthors}%
Mousavinasab, E.%
, Zarifsanaiey, N.%
, R.~Niakan~Kalhori, S.%
, Rakhshan, M.%
, Keikha, L.%
\BCBL {}\ \BBA {} Ghazi~Saeedi, M.%
\end{APACrefauthors}%
\unskip\
\newblock
\APACrefYearMonthDay{2021}{}{}.
\newblock
{\BBOQ}\APACrefatitle {Intelligent tutoring systems: a systematic review of characteristics, applications, and evaluation methods} {Intelligent tutoring systems: a systematic review of characteristics, applications, and evaluation methods}.{\BBCQ}
\newblock
\APACjournalVolNumPages{Interactive Learning Environments}{29}{1}{142--163}.
\PrintBackRefs{\CurrentBib}

\bibitem [\protect \citeauthoryear {%
Pane%
, Steiner%
, Baird%
\BCBL {}\ \BBA {} Hamilton%
}{%
Pane%
\ \protect \BOthers {.}}{%
{\protect \APACyear {2015}}%
}]{%
pane2015continued}
\APACinsertmetastar {%
pane2015continued}%
\begin{APACrefauthors}%
Pane, J\BPBI F.%
, Steiner, E\BPBI D.%
, Baird, M\BPBI D.%
\BCBL {}\ \BBA {} Hamilton, L\BPBI S.%
\end{APACrefauthors}%
\unskip\
\newblock
\APACrefYearMonthDay{2015}{}{}.
\newblock
{\BBOQ}\APACrefatitle {Continued Progress: Promising Evidence on Personalized Learning.} {Continued progress: Promising evidence on personalized learning.}{\BBCQ}
\newblock
\APACjournalVolNumPages{Rand Corporation}{}{}{}.
\PrintBackRefs{\CurrentBib}

\bibitem [\protect \citeauthoryear {%
Raes%
\ \protect \BOthers {.}}{%
Raes%
\ \protect \BOthers {.}}{%
{\protect \APACyear {2020}}%
}]{%
raes2020learning}
\APACinsertmetastar {%
raes2020learning}%
\begin{APACrefauthors}%
Raes, A.%
, Vanneste, P.%
, Pieters, M.%
, Windey, I.%
, Van Den~Noortgate, W.%
\BCBL {}\ \BBA {} Depaepe, F.%
\end{APACrefauthors}%
\unskip\
\newblock
\APACrefYearMonthDay{2020}{}{}.
\newblock
{\BBOQ}\APACrefatitle {Learning and instruction in the hybrid virtual classroom: An investigation of students’ engagement and the effect of quizzes} {Learning and instruction in the hybrid virtual classroom: An investigation of students’ engagement and the effect of quizzes}.{\BBCQ}
\newblock
\APACjournalVolNumPages{Computers \& Education}{143}{}{103682}.
\PrintBackRefs{\CurrentBib}

\bibitem [\protect \citeauthoryear {%
Rajalingam%
, Kanagamalliga%
, Karuppiah%
\BCBL {}\ \BBA {} Caesar~Puoza%
}{%
Rajalingam%
\ \protect \BOthers {.}}{%
{\protect \APACyear {2021}}%
}]{%
rajalingam2021peer}
\APACinsertmetastar {%
rajalingam2021peer}%
\begin{APACrefauthors}%
Rajalingam, S.%
, Kanagamalliga, S.%
, Karuppiah, N.%
\BCBL {}\ \BBA {} Caesar~Puoza, J.%
\end{APACrefauthors}%
\unskip\
\newblock
\APACrefYearMonthDay{2021}{}{}.
\newblock
{\BBOQ}\APACrefatitle {Peer interaction teaching-learning approaches for effective engagement of students in virtual classroom} {Peer interaction teaching-learning approaches for effective engagement of students in virtual classroom}.{\BBCQ}
\newblock
\APACjournalVolNumPages{Journal of Engineering Education Transformations}{34}{Special Issue}{425--432}.
\PrintBackRefs{\CurrentBib}

\bibitem [\protect \citeauthoryear {%
Reeve%
}{%
Reeve%
}{%
{\protect \APACyear {2012}}%
}]{%
reeve2012self}
\APACinsertmetastar {%
reeve2012self}%
\begin{APACrefauthors}%
Reeve, J.%
\end{APACrefauthors}%
\unskip\
\newblock
\APACrefYearMonthDay{2012}{}{}.
\newblock
{\BBOQ}\APACrefatitle {A self-determination theory perspective on student engagement} {A self-determination theory perspective on student engagement}.{\BBCQ}
\newblock
\BIn{} \APACrefbtitle {Handbook of research on student engagement} {Handbook of research on student engagement}\ (\BPGS\ 149--172).
\newblock
\APACaddressPublisher{}{Springer}.
\PrintBackRefs{\CurrentBib}

\bibitem [\protect \citeauthoryear {%
Roschelle%
, Penuel%
\BCBL {}\ \BBA {} Shechtman%
}{%
Roschelle%
\ \protect \BOthers {.}}{%
{\protect \APACyear {2006}}%
}]{%
roschelle2006co}
\APACinsertmetastar {%
roschelle2006co}%
\begin{APACrefauthors}%
Roschelle, J.%
, Penuel, W.%
\BCBL {}\ \BBA {} Shechtman, N.%
\end{APACrefauthors}%
\unskip\
\newblock
\APACrefYearMonthDay{2006}{}{}.
\newblock
{\BBOQ}\APACrefatitle {Co-design of innovations with teachers: Definition and dynamics} {Co-design of innovations with teachers: Definition and dynamics}.{\BBCQ}
\newblock

\PrintBackRefs{\CurrentBib}

\bibitem [\protect \citeauthoryear {%
Shemshack%
\ \BBA {} Spector%
}{%
Shemshack%
\ \BBA {} Spector%
}{%
{\protect \APACyear {2020}}%
}]{%
shemshack2020systematic}
\APACinsertmetastar {%
shemshack2020systematic}%
\begin{APACrefauthors}%
Shemshack, A.%
\BCBT {}\ \BBA {} Spector, J\BPBI M.%
\end{APACrefauthors}%
\unskip\
\newblock
\APACrefYearMonthDay{2020}{}{}.
\newblock
{\BBOQ}\APACrefatitle {A systematic literature review of personalized learning terms} {A systematic literature review of personalized learning terms}.{\BBCQ}
\newblock
\APACjournalVolNumPages{Smart Learning Environments}{7}{1}{33}.
\PrintBackRefs{\CurrentBib}

\bibitem [\protect \citeauthoryear {%
Sol{\'e}-Beteta%
, Navarro%
, Gaj{\v{s}}ek%
, Guadagni%
\BCBL {}\ \BBA {} Zaballos%
}{%
Sol{\'e}-Beteta%
\ \protect \BOthers {.}}{%
{\protect \APACyear {2022}}%
}]{%
sole2022data}
\APACinsertmetastar {%
sole2022data}%
\begin{APACrefauthors}%
Sol{\'e}-Beteta, X.%
, Navarro, J.%
, Gaj{\v{s}}ek, B.%
, Guadagni, A.%
\BCBL {}\ \BBA {} Zaballos, A.%
\end{APACrefauthors}%
\unskip\
\newblock
\APACrefYearMonthDay{2022}{}{}.
\newblock
{\BBOQ}\APACrefatitle {A data-driven approach to quantify and measure students’ engagement in synchronous virtual learning environments} {A data-driven approach to quantify and measure students’ engagement in synchronous virtual learning environments}.{\BBCQ}
\newblock
\APACjournalVolNumPages{Sensors}{22}{9}{3294}.
\PrintBackRefs{\CurrentBib}

\bibitem [\protect \citeauthoryear {%
Wei%
\ \BBA {} Jia%
}{%
Wei%
\ \BBA {} Jia%
}{%
{\protect \APACyear {2021}}%
}]{%
wei2021review}
\APACinsertmetastar {%
wei2021review}%
\begin{APACrefauthors}%
Wei, X.%
\BCBT {}\ \BBA {} Jia, H.%
\end{APACrefauthors}%
\unskip\
\newblock
\APACrefYearMonthDay{2021}{}{}.
\newblock
{\BBOQ}\APACrefatitle {A Review of the Application of Artificial Intelligence in the Virtual Learning Environment} {A review of the application of artificial intelligence in the virtual learning environment}.{\BBCQ}
\newblock
\BIn{} \APACrefbtitle {2021 Tenth International Conference of Educational Innovation through Technology (EITT)} {2021 tenth international conference of educational innovation through technology (eitt)}\ (\BPGS\ 79--82).
\PrintBackRefs{\CurrentBib}

\end{thebibliography}





\end{document}